\documentclass[]{spie}  

\newcommand{\nbar}{\bar{n}}

\usepackage{amsmath,amsfonts,amssymb}
\usepackage{graphicx}
\usepackage[colorlinks=true, allcolors=blue]{hyperref}
\usepackage{braket}
\title{Quantum-Assisted Optical Interferometers: Instrument Requirements}

\author[a]{Andrei Nomerotski}
\author[a]{Paul Stankus}
\author[a]{An\v{z}e Slosar}
\author[b]{Stephen Vintskevich}
\author[a,c]{Shane Andrewski}
\author[a]{Gabriella Carini}
\author[a]{Denis Dolzhenko}
\author[d]{Duncan England}
\author[c]{Eden Figueroa}
\author[c]{Sonali Gera}
\author[a]{Justine Haupt}
\author[a]{Sven Herrmann}
\author[a]{Dimitrios Katramatos}
\author[a]{Michael Keach}
\author[a]{Alexander Parsells}
\author[a]{Olli Saira}
\author[a]{Jonathan Schiff}
\author[e]{Peter Svihra}
\author[a]{Thomas Tsang}
\author[d]{Yingwen Zhang}
\affil[a]{Brookhaven National Laboratory, Upton NY 11973, USA}
\affil[b]{Moscow Institute of Physics and Technology,
Dolgoprudny, Moscow Region 141700, Russia}
\affil[c]{Department of Physics and Astronomy, Stony Brook University, Stony Brook, NY 11794, USA}
\affil[d]{National Research Council of Canada, 100 Sussex Drive, Ottawa, Ontario, K1A 0R6, Canada}
\affil[e]{Department of Physics and Astronomy, School of Natural Sciences, 
University of Manchester, Manchester M13 9PL, United Kingdom}

\authorinfo{Further author information: E-mail: anomerotski@bnl.gov, Telephone: 1 631 3448338}

\begin{document} 
\maketitle

\begin{abstract}
It has been recently suggested that optical interferometers may not require a phase-stable optical link between the stations if instead sources of quantum-mechanically entangled pairs could be provided to them, enabling extra-long baselines and benefiting numerous topics in astrophysics and cosmology. We developed a new variation of this idea, proposing that  photons from two different sources could be interfered at two decoupled stations, requiring only a slow classical connection between them.  We show that this approach could allow  high-precision measurements of the relative astrometry of the two sources, with a simple estimate giving angular resolution of $10 \  \mu$as in a few hours' observation of two bright stars. We also give requirements on the instrument for these observations, in particular on its temporal and spectral resolution. Finally, we discuss possible technologies for the instrument implementation and first proof-of-principle experiments.
\end{abstract}


\section{Introduction}
\label{sec:introduction}

Improving angular resolution in astronomical instrumentation has long been pursued in order to advance astrometric observation capabilities. In traditional telescopes, the resolution is diffraction-limited, given by the relationship $\Delta\theta \sim \lambda/D$ where $\lambda$ is the photon wavelength and $D$ is the aperture width. 
However, enlarging the aperture size is not the only technique at our disposal to improve angular resolution. Drawing on the trailblazing work of Michelson from the end of the 19th century, interferometers utilize classical interference effects to significantly improve  the angular resolution of telescopes. In the traditional interferometer, a single photon, which in its simplest form can be considered a plane wave, is detected by two telescopes at different detection sites. When the detected light is recombined and interfered, the emergent fringe pattern is sensitive to the phase difference incurred due to differences in the photon’s path length to each detection site. Interferometry uses telescopic networks to effectively create a telescope with an aperture size given by the distance separating the detection sites in the network. The angular resolution now scales as follows: $\Delta\theta \sim \lambda/B$ where $B$ is the baseline separation between the two detection sites, a dramatic improvement on single telescope detection methods.

Although traditional interferometry is responsible for some of the greatest modern observational feats, the required optical connective path between detection sites to interfere single photons limits the sensitivity of the traditional optical interferometer. In contrast, for radio-range wavelengths the signal amplitude can be measured directly and, therefore, the interference can be conducted offline after observation. Therefore, for radio interferometers, the baseline separation has successfully and cost-effectively been extended to thousands of kilometers in a technique aptly named Very Long Baseline Interferometry (VLBI). However, the cost of building and sustaining an optical path to achieve the improved resolution that a long-baseline optical interferometer provides has thus far restricted optical interferometers to a baseline distance of only hundreds of meters. 


In order to achieve the improved resolution that an optical interferometer can provide, this paper discusses the instrument requirements for a novel type of optical interferometer first proposed in our previous work \cite{stankus2020} that utilizes quantum mechanical interference effects between two photons from two astronomical sources. This development can be considered as a practical variation of pioneering ideas described in the work of Gottesman, Jennewein and Croke in 2012 \cite{Gottesman2012} to employ a source of entangled photon pairs to measure the photon phase difference between the receiving stations, which were further developed in the following references \cite{harvard1, harvard2}.

By using the two sky sources the proposed interferometer bypasses the traditional necessity of establishing a live optical path connecting detection sites. The freedom to utilize traditional telecommunication methods to relay information between detection sites and extract quantum mechanical interference effects allows for the drastic resolution improvement that an optical interferometer provides without the restrictive and costly optical path that traditional interferometric techniques necessitate. 
In the proposed two-photon amplitude interferometer, the baseline distance can be made arbitrarily large, and consequently, an improvement of several orders of magnitude in angular resolution is attainable. Although a full quantum field theory treatment, as is found in our previous work\cite{stankus2020}, is necessary to rigorously describe the interference effects the two-photon interferometer utilizes, we will limit ourselves in this paper to a quantum optics based description, which although incomplete, is still illustrative of the theory behind the instrument.

The operation of the proposed two-photon interferometer imposes imaging and spectroscopic sensitivity requirements. Fast imaging with sufficiently good temporal resolution and fine spectral binning is needed in order to measure  single-photon events at each detection site and perform the necessary correlative analysis. We will therefore devote the second half of this paper to an exploration of some of the currently available technologies that meet these sensitivity criteria and discuss our preliminary results from their usage in tests with spontaneous parametric down conversion (SPDC) sources and pseudo-thermal argon lamps.

The rest of the paper is organized as follows: Section 2 highlights cosmological and astrophysical topics that could benefit from improved astrometric precision. Section 3 discusses a quantum optics based description of the two-photon amplitude interferometry and new observables based on the measurement of the Earth rotation fringe rate. Section 4 discusses the instrumentational requirements that must be met to satisfy needs  of  the  two-photon interferometer. Sections 5  and 6 discuss currently available  spectroscopic binning techniques and fast imaging technologies that meet these requisite sensitivity criteria.  Lastly, section 7 describes the experimental setup and preliminary results from proof-of-concept lab testing.

\section{Science Applications}

It is difficult to compile an exhaustive list of all of the astrophysical and cosmological applications of the proposed optical interferometer. We will therefore briefly touch on just a few of the many scientific opportunities afforded by considerable improvements in astrometric precision. 

\textbf{Testing theories of gravity by direct imaging of black hole accretion discs}: It has long been recognized that the extreme gravitational effects that take place at the surface of black holes provide a unique testing ground for modified theories of gravity. An optical array with a baseline distance comparable to the Event Horizon Telescope (EHT) would provide significant improvement on the already impressive 25 microarcsecond resolution achieved by the EHT’s Earth-spanning radio telescopic network, which generated the now famous image of the black hole event horizon in M87 \cite{2019ApJ...875L...2E}. This would allow for more detailed imaging of the scientifically rich accretion disks on the event horizon surrounding black holes and would provide new opportunities to better understand these enigmatic regions.

\textbf{Precision parallax and cosmic distance ladder}: The  discrepancy between measurements of the expansion rate of the Universe that rely on distant Type Ia supernovae and those based on redshift measurements of baryonic acoustic oscillations and the cosmic microwave background is currently one of the most vexing inconsistencies in observational cosmology. Since we cannot directly measure the distance to Type Ia supernovae in distant galaxies, we rely on a multi-step process in which local measurements are used to extract the distance to Type Ia supernovae at cosmological scales, a process commonly referred to as the cosmic distance ladder. Significant improvements in astrometric precision would allow for direct parallax measurements of galaxies with Type Ia supernovae. In so doing, we could skip over several presently unavoidable rungs of the distance ladder, improving its accuracy by removing a few of its potentially error-inducing steps. 

\textbf{Mapping microlensing events}: Better understanding the nature of dark matter (DM) occupies a large portion of contemporary physics research. Amongst the candidates for DM are compact star-sized objects, such as black holes, or extended virialized subhalos comprised of yet undiscovered dark matter particles. To probe these DM candidates, a more rigorous and direct method of observing their expected gravitational microlensing effects on stars is needed. The typical photometric measurement of microlensing events both obfuscates details of the lens’s mass and spatial distribution and is less straightforward than an astrometric approach. Improvements in astrometric precision would allow for the more direct astrometric approach to mapping microlensing effects and would therefore be beneficial in assessing the viability of certain DM candidates.

\textbf{Peculiar motions and dark matter}: DM’s effects on the dynamics of our Galaxy are of great interest for understanding its properties and local distribution. The ability to fully measure and reconstruct 3D velocities of a large swath of the stars in the Galaxy would unlock important, thus far inaccessible data that could illuminate many of the unknown characteristics of DM in our galaxy. Improvements in astrometric measurements are needed to measure the peculiar motion of more distant stars in our Galaxy and subsequently extract their transverse velocity. The improved 3D velocity data afforded by more precise astrometric measurements would pave the way for an inferred measurement of the dark matter halo’s gravitational potential. Moreover, it would allow us to survey historical merging events in the Milky Way halo and open a unique window into DM’s interaction with itself and with ordinary matter \cite{Chu2019}. 

\section{Two-photon amplitude interferometry}
\label{sec:basics_two_photon_amplitude}

Classical single-photon interferometry involves detecting a single source photon at multiple stations. On the other hand, the main focus of this paper is two-photon interferometry, a novel technique where two photons from two different sources are interfered.

In our previous work, we used a quantum description of interferometry, which we performed in two stages \cite{stankus2020}. The first used a more intuitive quantum mechanical picture of monochromatic photons as particles; e.g. definite Fock states, carried forward in a Schr\"{o}dinger representation.
The second approach employed a full quantum field theory calculation with time-dependent electric field operators.  These calculations allowed us to properly address the quasi-monochromatic case and the time correlations between the measurements of the two photons, as well as extended sources. Both descriptions provided consistent results, as they should.

In the below section, we give a short theoretical summary of the observable two-photon interference based on the standard formalism of quantum optics, closely following the method described in our prior work \cite{stankus2020}.

\subsection{Observation of two-photon amplitude interference}

The radiation from two distant stars is believed to be thermal and it is assumed that the photon occupation numbers are small for each quantum-mechanical mode. One can describe the field propagation in terms of electric field operators where the information about angular distribution is encoded in phases. For faraway objects like stars or galaxies, the far-field approximation can be used, so the approach is closely related to the Van Zitter-Zernike Theorem, and, therefore, methods of Fourier optics \cite{goodman2005,mandel_wolf1995} could provide a solid theoretical description of the problem. Below we also assume that the quasi-monochromatic approximation is valid since we use spectral binning in the observation scheme. Based on these simple assumptions, let us model a positive frequency part of the electric field operator within one spectral bin as follows:

\begin{eqnarray}\label{propagation}
\hat{E}^{(+)}_{j} =
\int \tilde{G}^{[j]}_{\vec{R}_{j}}(\omega,\vec{k}_{\perp})\hat{a}^{[j]}(\vec{k}_{\perp},\omega)e^{i\omega (\frac{R_{j}}{c} - t+\delta_{js})} d\vec{k}_{\perp}d\omega,\ \ j=1,2 ;
\end{eqnarray}

where we introduced the function:

\begin{eqnarray}\label{simplified-g-a}
\tilde{G}^{[j]}_{\vec{R}_{js}}(\omega,\vec{k}_{\perp}) = \sqrt{\frac{\hbar \omega^3}{(2 \pi)^3 R_{js}^2 c^2}} 
\int_{\Sigma_{j}}e^{i\left(\vec{k}_{\perp} - \frac{\omega}{c}\frac{\vec{R}_{js}}{R_{js}}\right)\vec{r}_{0\perp}^{(j)}}e^{i\frac{\omega\vec{r}^{2}_{0\perp}}{2cR_{j}}}d\vec{r}_{0\perp}^{(j)}, \ \ j=1,2.
\end{eqnarray}

\begin{figure}
\begin{center}
\includegraphics[width=1\linewidth]{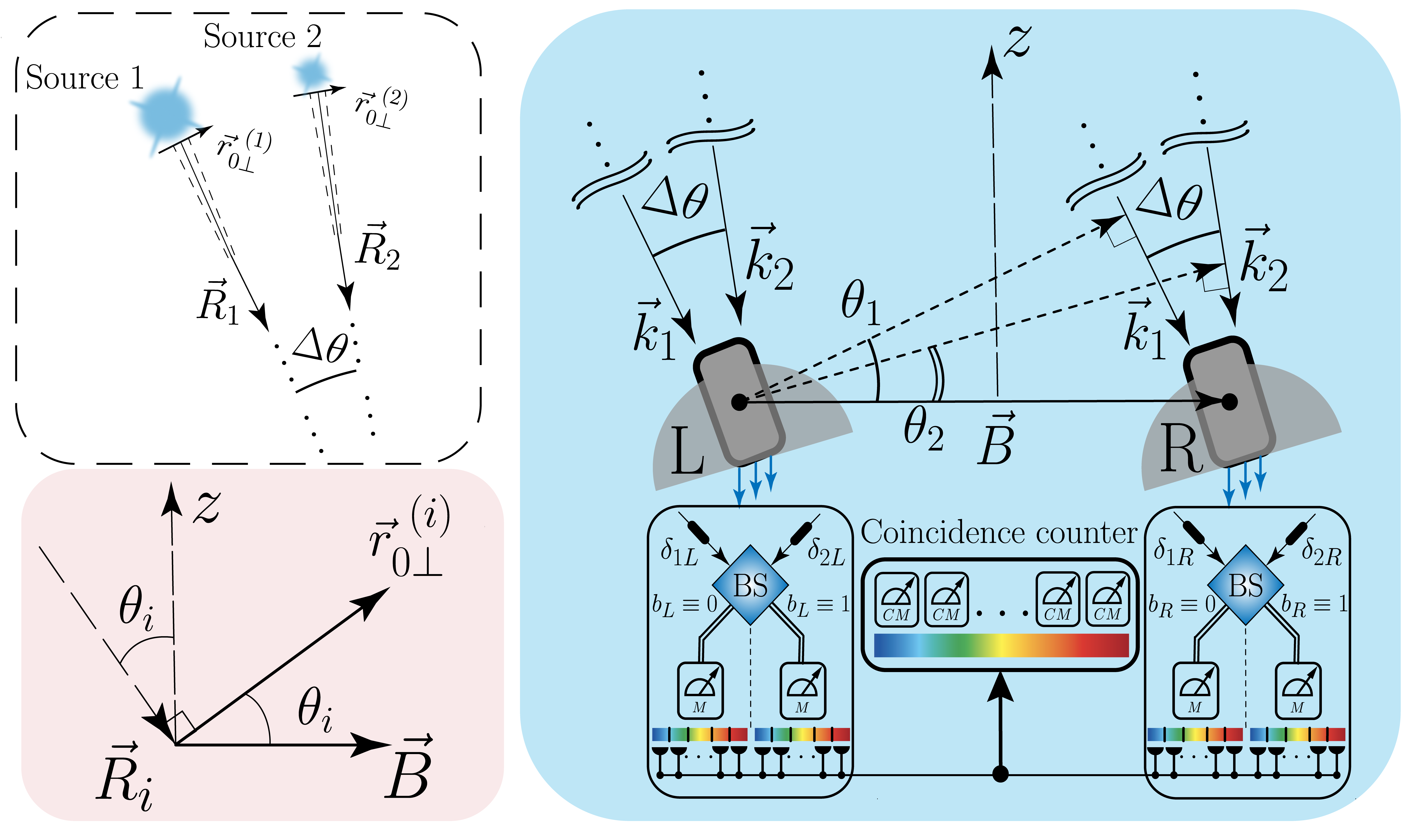}
\caption{
Geometry of the observation scheme. In the transparent (top left) sub-panel, vectors $\vec{R}_{1}$ and $\vec{R}_{2}$ are collinear to wave-vectors $\vec{k}_1$ and $\vec{k}_2$ respectively and indicate the direction of incident wave vectors from two thermal sources (stars) within the far-field approximation. $\Delta \theta $ is an opening angle.
The pink (bottom left) sub-panel illustrates the mutual arrangement of vectors presented in derivations. Angles $\theta_{1}$ and $\theta_{2}$ are equatorial polar angles of sources relative to the axis defined by the vector $z$, a direction to the zenith which is orthogonal to the baseline vector $\vec{B}$. For a given direction of $\vec{R}_{i}$ with $i=1,2$ one can find the orthogonal projection of $\vec{B}$ onto the plane of the corresponding source, where each point of this plane is specified by a vector $\vec{r}_{0\perp}^{i}$. In the blue (right) sub-panel the radiation from both sources is collected at observation stations marked by letters $L$ and $R$ and mixed in the beam-splitters (BS). Output channels of BS at each station are labelled by binary variables $b_{L}$ and $b_{R}$ respectively, and we depict phase delays for each input port of the BS by $\delta_{js}, \ j=1,2 \ s = L,R$. We assume that the radiation spectrum at each output port is divided into narrow spectral bins of width $\Delta \omega $. 
For each spectral bin, we analyze coincidence events between separate detectors, one at station $L$ and one at $R$, to observe two-photon interference and extract  information about the opening angle $\Delta \theta$, see Equation \eqref{coince_2}.}
\label{geometry}
\end{center}
\end{figure}

We denoted  $j = {1,2}$ as the index of a specific source and $\Sigma_{j}$ as a characteristic area of the source projection on the object plane where each point can be considered as an independent sub-source. We consider contributions from each independent sub-source by integrating over the $d\vec{r}_{0\perp}$ -coordinate in the source plane; $R_{j}$ is a distance between the $j$-th source and observation point, and $\delta_{js}$ is an additional and adjustable phase delay which occurs due to additional paths introduced at each observation station, marked as $L$ and $R$, see Figure \ref{geometry}.
Note that to simplify the overall description we assumed that polarization is fixed to the same value for all photons.

One may assume that optical instruments which at each observation station can collect radiation from two independent sources $1$ and $2$, resolving two independent spatial modes $\vec{k}_{1}$ and $\vec{k}_{2}$ respectively. Both modes are collected at each station and fed to the beam-splitter~(BS) input port, as in Figure \ref{geometry}. As a result the superimposed field after the BS transformation can be written as follows:

\begin{eqnarray}\label{field_operator_init}
\hat{E}^{(+)}_{b_{s}} = \frac{1}{\sqrt{2}}\left(\hat{E}^{{[1]}{(+)}}_{s} + (-1)^{b_{s}}\hat{E}^{{[2]}{(+)}}_{s}\right); \ 
b_{s}\in \{0,1\}, \ s\in \{L,R\}; 
\end{eqnarray}

where the binary index $b_{s}$ points to the BS's specified output port for each observation station: left with $s = L$ or right with $s = R$. 
One can characterize each spectral bin by the central frequency $\omega_{0}$ and specific spectral bandwidth $\Delta\omega$, assuming for simplicity that our photodetector quantum efficiency is almost independent of frequency. We can model the joint probability of registration between detectors at each station through the fourth-order coherence function. In accordance with the standard photodetection theory \cite{Glauber1963}, the function can be expressed as:

\begin{eqnarray}\label{corr.avarage}
&&\Gamma_{1,2}^{b_{L},b_{R}} = \braket{\hat{E}^{{(-)}}_{b_{L}}\hat{E}^{{(-)}}_{b_{R}}\hat{E}^{{(+)}}_{b_{R}}\hat{E}^{{(+)}}_{b_{L}}} =  \braket{\mathcal{T}:\hat{I}^{[1]}_{L}\hat{I}^{[1]}_{R}:}+\braket{\mathcal{T}:\hat{I}^{[2]}_{L}\hat{I}^{[2]}_{R}:}+\braket{\mathcal{T}:\hat{I}^{[1]}_{L}\hat{I}^{[2]}_{R}:}+\braket{\mathcal{T}:\hat{I}^{[2]}_{L}\hat{I}^{[1]}_{R}:}+ \nonumber \\
&&\left(-1\right)^{b_{L}+b_{R}}\bigg[\braket{\mathcal{T}\hat{E}^{{[1]}{(-)}}_{R}\hat{E}^{{[2]}{(-)}}_{L}\hat{E}^{{[2]}{(+)}}_{R}\hat{E}^{{[1]}{(+)}}_{L}} + {\rm{c.c.}}\bigg],
\end{eqnarray}

where the angle brackets $\braket{}$ indicate that all observables are averaged over an ensemble of multimode thermal states, \cite{mandel_wolf1995,Glauber1963} taking also into account randomness of phases in thermal sources and assuming a cross-spectral purity for stationary optical fields radiated by each source \cite{mandel_wolf1995}. We also employed the intensity operator $\hat{I}^{j}_{s}= \hat{E}^{[j](-)}_{s}\hat{E}^{[j](+)}_{s}$ with $s \in \{L,R\}$ and $j\in\{1,2\}$, while the symbols $\mathcal{T}$ and $::$  indicate that all operators inside expressions like  $\braket{\mathcal{T}:_{\dots}:}$  must be time and normal ordered~\cite{mandel_wolf1995,Ou1988,Mandel1983}. The result of substitution of \eqref{propagation}, \eqref{simplified-g-a}, and \eqref{field_operator_init} in \eqref{corr.avarage} results in the following explicit form of the fourth-order coherence function:

\begin{eqnarray}\label{corr.rewritten}
&&\Gamma_{1,2}^{a_{l},a_{r}} \approx I_{1}^2(1+|\gamma_{1}(\omega_{0})|^2) + I_{2}^2(1+|\gamma_{2}(\omega_{0})|^2) +2I_{1}I_{2} \times \nonumber \\ 
&&\bigg[1 + (-1)^{(a_{L}+a_{r})}|\gamma_{1}(\omega_{0})||\gamma_{2}(\omega_{0})|\times \cos\left(\frac{\omega_{0}b}{c}(\sin\theta_{1} - \sin\theta_{2}) + \frac{\omega_{0}\Delta L}{c}\right)\bigg], 
\end{eqnarray}

where $I_{1}$ and $I_{2}$ are the average intensities of the sources, for simplicity assuming narrow spectral filtering $\Delta\omega \longrightarrow 0$. Thus, we can define
$\gamma_{j}(\omega_{0})$, $j=1,2$ as follows:

\begin{eqnarray}
\gamma_{j}(\omega_{0}) = \frac{\braket{\hat{E}^{{[j]}{(-)}}_{L}\hat{E}^{{[j]}{(+)}}_{R}}}{I_{j}} \approx 
\frac{1}{\Sigma_{j}}{\rm{FT}}_{\Sigma_{j}}\left(\frac{\omega\vec{B}\cdot\vec{r}_{0\perp}^{j}}{c}\right)
e^{i\omega_{0}\left(\frac{B\sin(\theta_{j})}{c} - \tau+\delta_{jR} -\delta_{jL}\right)}, \ \Delta\omega \longrightarrow 0;
\end{eqnarray}

where we set the phase delays as $(\delta_{1R} - \delta_{1L} - \delta_{2R} + \delta_{2L})/c \equiv \Delta L/c$. We also denoted $\frac{1}{\Sigma_{j}}{\rm{FT}}_{\Sigma_{j}}\left(\frac{\omega\vec{B}\cdot\vec{r}_{0\perp}^{j}}{c}\right)$ as the normalized Fourier coefficient of the source distribution in the case of the sharp-edge disc model for each source with effective diameter $D_{j}$ with  $\Sigma_{j} = \pi D_{j}^2/4$. Thus, the normalized Fourier coefficient is given by:

\begin{eqnarray}\label{fourier-trans}
\frac{1}{\Sigma_{j}}{\rm{FT}}_{\Sigma_{j}}\left(\frac{\omega\vec{B}\cdot\vec{r}_{0\perp}^{j}}{c}\right) =  
\frac{ 2J_{1}\left(\frac{\omega_{0}}{2c R_{j}}BD_{j}\cos{\theta_{1}}\right)}{\frac{\omega_{0}}{2c R_{j}}BD_{j}\cos{\theta_{j}}}, j=1,2;
\end{eqnarray}

where $J_{1}(x)$ is a Bessel function of the first kind. We refer to our recent paper \cite{stankus2020} for a detailed derivation of the explicit forms of Equations \eqref{corr.avarage} and \eqref{corr.rewritten}. Finally, the observable photon coincidence rate is proportional to fourth-order coherence \eqref{corr.rewritten}:

\begin{eqnarray}\label{coince_2}
N_{c}(b_{L},b_{R})\propto \big[1 + (-1)^{(b_{L}+b_{R})} V \cos\left(\frac{\omega_{0}B(\sin\theta_{1} - \sin\theta_{2})}{c} + \frac{\omega_{0}\Delta L}{c}\right)\big].
\end{eqnarray}

The quantity $V$ can be considered as a fringe visibility, which in the case of very narrow frequency filter with $\Delta \omega \rightarrow 0$ can be written as follows: 

\begin{eqnarray}\label{Vis}
V =
\frac{2I_{1}I_{2}\xi_{1}\xi_{2}}{(I_{1}+I_{2})^2 + (I_{1}\xi_{1})^2 + (I_{2}\xi_{2})^2},
\end{eqnarray}

where we used $\xi_{j} \equiv  \frac{1}{\Sigma_{j}}{\rm{FT}}_{\Sigma_{j}}\left(\frac{\omega\vec{B}\cdot\vec{r}_{0\perp}^{j}}{c}\right)$ to simplify the expression.
However, in real experiments the bandwidth $\Delta\omega$ is finite, as is the coherence time $\tau_{c} \propto \frac{1}{\Delta\omega}$, which may decrease the visibility for observation of two-photon correlation effects in the presented scheme. Of course, the final coincidence rate will depend on the detector quantum  efficiency, collecting area and the ratio between resolution time and coherence time.

We recognize that this two-photon technique has similarities with the  Hanbury Brown \& Twiss~(HBT) intensity correlation technique~\cite{hbt, Foellmi2009}, which also performed optical interferometry using two independent stations connected with only a slow classical link. This is apparent in Equation (5), which can be decomposed into three parts corresponding to (i) random coincidences; (ii) HBT and (iii) new, phase-dependent term. We note that the HBT scheme does not allow one to reconstruct the phase information.
In contrast, in the presented approach, transformation of the photon states at the beam-splitters allows one to extract the angular information encoded in optical paths with higher accuracy. This is only possible because we are able to distinguish the spatial modes from each source. Thus, the proposed two-photon amplitude interferometry captures more information from the photon field and, essentially, can be seen as a generalization of the original HBT technique.

\subsection{Earth rotation fringe rate observable}
\label{subsec:drift_scan_fringe_rate}

In our prior work \cite{stankus2020} we proposed to use the Earth rotation fringe rate as a convenient observable that can be measured with good precision to determine the opening angle between two sky sources. In this case, we can write the average number of observed pair coincidences as a function of time with four parameters:
\begin{equation}
    \langle N({b_{L}b_{R}}) \rangle (t) =
    \bar{N}({b_{L}b_{R}}) \left[ 1 \pm  V \cos \left(\omega_{f} t + \Phi \right) \right].
    \label{eq:basic_time_dependence}
\end{equation}
\noindent
Here we use $\bar{N}({b_{L}b_{R}})$ for the average observed number of pairs of type $({b_{L}b_{R}})$, with the ``$+$'' and ``$-$'' corresponding to the different pair types; e.g. $b_{L} \equiv 0, b_{R} \equiv 0$, $b_{L} \equiv 0,b_{R} \equiv 1$ versus $b_{L} \equiv 0, b_{R} \equiv 1$, $b_{L} \equiv 1, b_{R} \equiv 0$;  $V$ is the fringe visibility defined earlier;
and $\Phi$ is an overall phase offset reflecting the delays in the system and due to the sources sky positions. The fringe angular rate $\omega_{f}$ is 
\begin{equation}
\omega_{f} = \frac{2 \pi B \Omega}{\lambda} 
(\sin \theta_{0} \sin \Delta\theta + \cos \theta_{0}
  (1-\cos\Delta\theta));
  \label{eq:nu_fringe_full}
\end{equation}
which provides a direct measure of $\Delta\theta$ if all the other parameters are known. $\theta_{0}$ is the position of source 1 at the epoch chosen as $t=0$, $\Delta \theta$ is the opening angle between the sources, and $\Omega=1.16\times$10$^{-5}$ rad/sec is the angular velocity of the Earth's rotation. 

In the limit of small opening angle $\Delta\theta \ll 1$ the fringe rate simplifies to
\begin{equation}
\omega_{f} = \frac{2 \pi B \Omega \sin \theta_{0}  }{\lambda} \Delta\theta .
  \label{eq:nu_fringe_smallopen}
\end{equation}

\subsection{Astrometric precision}

Following our previous work, \cite{stankus2020} we estimate the relative precision on the Earth rotation fringe rate using the Fisher matrix formalism, which provides the expected sensitivity for optimal estimators:

\begin{equation}
    \frac{\sigma \left[ \omega_{f} \right]}{\omega_{f}} =
    \frac{\sigma \left[ \Delta \theta \right]}{\Delta\theta} =
\sqrt{\frac{6}{\pi^{2} \kappa}} \; \frac{1}{V \, N^{Cycle} \sqrt{N^{Pair}}}
\label{eq:stdv_dimensionless}
\end{equation}

where $\kappa(V)$ is a small dimensionless auxiliary function with a value between $1/2$ and $1$. 
As can be seen, the fractional uncertainty on $\omega_{f}$ and on $\Delta\theta$ depends inversely on the three dimensionless quantities: (i)~the fringe visibility $V$;
(ii)~the number of full fringe cycles $N^{Cycle}=T \omega_{f}/2\pi$ that pass during the observation time $T$; and (iii)~the square root of the total number of observed pairs $N^{Pair}=\nbar T$.  

Further, we can write the uncertainty on $\Delta\theta$ to determine our sensitivity to astrometric changes between observations on different days:

\begin{equation}
    \sigma \left[ \Delta \theta \right] =
\sqrt{\frac{6}{\pi^{2} \kappa}} \; \frac{1}{V} \, 
\frac{\lambda}{B} \, 
\frac{1}{ T \Omega \, \sin \theta_{0} } \,
\frac{1}{\sqrt{\nbar T}}
\label{eq:stdv_deltatheta}
\end{equation}

where $\nbar$ is the average rate of pairs from both sources.

We note the following two observations from Equation~\eqref{eq:stdv_deltatheta}: (i)~as expected, the uncertainty on $\Delta\theta$ scales with $\lambda/B$, allowing us to benefit from longer baselines as long as the visibility is adequate; and (ii)~the overall $T^{-3/2}$ dependence on the length of the observation period is considerably faster than the simple $T^{-1/2}$ gain from the photon coincidence statistics, reflecting the advantage of the rate measurement, in contrast to other observables.

Following an example of a pair of magnitude 2 stars \cite{stankus2020} registered with two telescopes with 1 m$^2$ effective diameter (which folds in the photon detection efficiency), 200 m baseline, 0.15 ns time detection binning and deploying $4 \cdot 10^4$ sets of spectral bins, we estimate a precision of the order of $\sigma [ \Delta\theta ]\sim10\ \mu$as for one night's observation.
A brief discussion of possible systematic errors such as those due to the atmosphere is given in our recent work \cite{stankus2020}.

\section{Instrument requirements}

In this section we will focus on the requirements of the telescope camera, which needs to perform fast spectroscopy of registered photons.

\begin{figure}
\begin{center}
\includegraphics[width=0.7\linewidth]{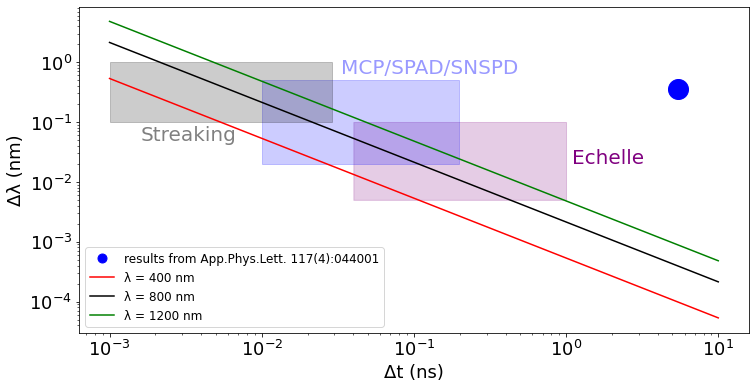}
\caption{Temporal and spectral resolution requirements for various wavelengths necessary to satisfy photon indistinguishability under the Heisenberg uncertainty principle formulation: $\Delta t \cdot \Delta \lambda = const$. The temporal resolution of streaking cameras, MCP, SPAD, and SNSPD imaging technologies, and the spectral resolution of different gratings sets a range of corresponding temporal and spectral resolutions, which are displayed as well. Blue dot shows the results achieved in our previous work \cite{Svihra2020}.}
\label{fig:uncertainty}
\end{center}
\end{figure}

An important conclusion of the previous section was that the photons must be close enough in time and frequency to efficiently interfere or, formulating it differently, to be indistinguishable within the Heisenberg uncertainty principle $ \Delta t \cdot \Delta E \sim \hbar $. If converted to $\Delta t \cdot \Delta \lambda $ variables, this expression would correspond to isolines of $\Delta t \cdot \Delta \lambda$, which are presented in Figure \ref{fig:uncertainty} for three values of $\lambda$. The figure also presents several  projections of performance for fast devices that can be used for these measurements, discussed in more detail in the next section.

Figure \ref{fig:uncertainty} is useful to set the scale for the timing and spectral resolution requirements for the instrument. Each point on the graph can be interpreted as a temporal and spectral binning which provides the maximum visibility for the two-photon amplitude interference fringes. It is clear that there exists a trade-off between the spectral and temporal resolutions, so a better time resolution alleviates requirements on the spectral resolution and vice versa.

We note that if the wavelength bin width is larger than the width determined from Figure \ref{fig:uncertainty}, then the visibility will be lower. However, as can be seen from the previous section, the statistical uncertainty of the Earth rotation fringe rate, which is the main observable in the experiments,  will depend ultimately only on the number of collected photon coincidences, which will scale with the size of the wavelength bins at a given time binning. Therefore, what matters for precision is the total number of spectroscopic bins.

Another important parameter for the imaging system is the photon detection efficiency, which needs to be as high as possible, since the two-photon coincidences used in the proposed approach have a quadratic dependence on the photon detection efficiency.

We discuss below the existing technologies which have promise to satisfy these requirements.

\section{Possible technologies for spectroscopic binning}

The spectroscopic binning can be implemented employing a traditional diffraction grating spectrometer
where incoming photons pass though a slit, are dispersed by a diffraction grating, and are then focused onto an array of single-photon detectors.  A possible scheme is presented in the left part of Figure \ref{fig:quantumtarget}. In our previous study,  \cite{Zhang2020} we used a down-conversion source of correlated photon pairs to characterize the spectral (and temporal) binning. The photons in a pair are produced simultaneously and are also anti-correlated in energy (and, therefore, in wavelength) since their energy must sum up to the pump photon energy. The dispersion in the time-energy product of this measurement (right part of Figure \ref{fig:quantumtarget}) was orders of magnitude above the Heisenberg limit, so significant improvements in spectral and temporal resolution are required. These results are also shown as a blue dot in Figure \ref{fig:uncertainty}. Details of possible fast detector systems are outlined in section~\ref{sec:Fast_imaging}; here we focus on the spectroscopic resolution.

The ultimate resolving power of a grating spectrometer is given by:
\begin{equation}
\label{eq:Resolution}
    R = \frac{\lambda}{\Delta\lambda} = knW_g,
\end{equation}
where $\Delta\lambda$ is the minimum separation between two lines that can be resolved, at mean wavelength $\lambda$. The resolving power is proportional to the refraction order $k$, the line density of the grating $n$, and the spot size at the grating $W_g$. Manufacturing considerations constrain the size and line density of a grating, so the resolving power cannot be increased arbitrarily. Furthermore, the resolution is also limited by a number of other factors, including the slit size, the focal length of the lens, and the pixel size of the detector, so $R$ can only be considered an upper bound. Applying realistic parameters to equation~\eqref{eq:Resolution} (1800\,g/mm grating, 10\,cm spot size), one can see that spectral resolution of around 0.05\,nm can be readily achieved with first-order diffraction. These simple first-order gratings use only a single stripe on the detector and so do not utilize the full 2-dimensional pixel array.

From equation~\eqref{eq:Resolution}, one can see that the resolving power depends upon the order of diffraction, so higher orders have stronger resolution. This is the basis of Echelle-type spectrometers in which a standard first-order grating is employed perpendicular to a second grating that is blazed to optimise multiple overlapping higher orders. High resolution is provided by the high-order diffraction, and the first-order grating vertically separates the overlapping modes. The output of an Echelle spectrometer is a series of parallel stripes, so the full 2-dimensional array is used to give high resolution across a large range. Typical resolution for this type of spectrometer is around 0.01\,nm~\cite{Vogt1994,Tull1995,Detalle2001}. It is interesting to note that spectral resolution could be improved to the sub-picometer scale by placing a Fabry-P\'erot etalon before the spectrometer~\cite{Deng2109}, though for a broadband light source like a star, this will reduce the collection efficiency of the spectrometer by approximately the finesse of the etalon. 

High-resolution spectroscopy is widely used in modern astronomical observations---for example, to allow precise measurement of redshift. Therefore it is likely that our proposed technique can ``piggy-back'' on existing astronomical technologies and, possibly, infrastructure simply by adding quantum detectors. To give examples of existing devices, the BOSS \cite{Dawson2012} 
and, more recently, DESI \cite{Dey2019} experiments have spectral resolution of about 0.2 nm across wide wavelength ranges. In the DESI case, the star light is spread using a volume phase holographic (VPH) grating over a CCD sensor with 4096×4096 15-micron square pixels. Three wavelength bands: blue, red and infrared cover the total range from 360 to 980 nm. Spectrometers with the Echelle principle can have much higher resolution; one example is the HIRES spectrograph \cite{Vogt1994} with resolution of about 0.01 nm, in operation on the Keck Telescope. It is based on a standard in-plane Echelle spectrometer with grating post-dispersion.

The higher spectral resolution almost invariably comes with a lower system throughput. The  DESI spectrograph throughput varies between 40\% and 75\%, while the peak efficiency of HIRES spectrograph is 13\%. Since our application is quadratically sensitive to system throughput, these losses can be significant. Any realistic experiment will thus have to optimize the parameters carefully for overall system performance, but this exceeds the scope of this contribution.

\section{Possible technologies for fast imaging systems}
\label{sec:Fast_imaging}

Fast imaging technologies are required in many fields of modern science, for example, to record rapid processes or measure time-of-flight and temporal coincidences. The so-called framing approach, where all information in the field-of-view is recorded, can be contrasted with the data-driven approach, where the empty pixels without useful information are suppressed. The latter has the advantage when the recorded data is sparse, since the zero suppression leads to substantial gain in the readout throughput and allows one to implement complex functionality at the pixel level, such as time stamping with nanosecond precision \cite{Nomerotski2019}, which is required for the above measurements of two-photon amplitude interferometry.

\subsection{Fast intensified optical cameras}

An example of a data-driven optical camera is Tpx3Cam \cite{tpx3cam, Nomerotski2019}. In this camera a light-sensitive, back-side illuminated silicon sensor is attached to the readout chip, Timepix3, \cite{timepix3} via bump-bonding. The camera has single-photon sensitivity after addition of an image intensifier, a vacuum device with a photocathode followed by a microchannel plate (MCP) and fast scintillator. 
The photon detection efficiency is determined by the photocathode and typically can achieve values of about 30\%.  The timing resolution achieved for single photons is 2 ns (rms) \cite{Ianzano2020}.
This camera was used for a variety of measurements with single photons including single photon counting \cite{Nomerotski2020}, characterization of single photon sources, \cite{Ianzano2020} and quantum target detection \cite{Svihra2020, Zhang2020}.

\begin{figure}
\begin{center}
\includegraphics[width=0.57\linewidth]{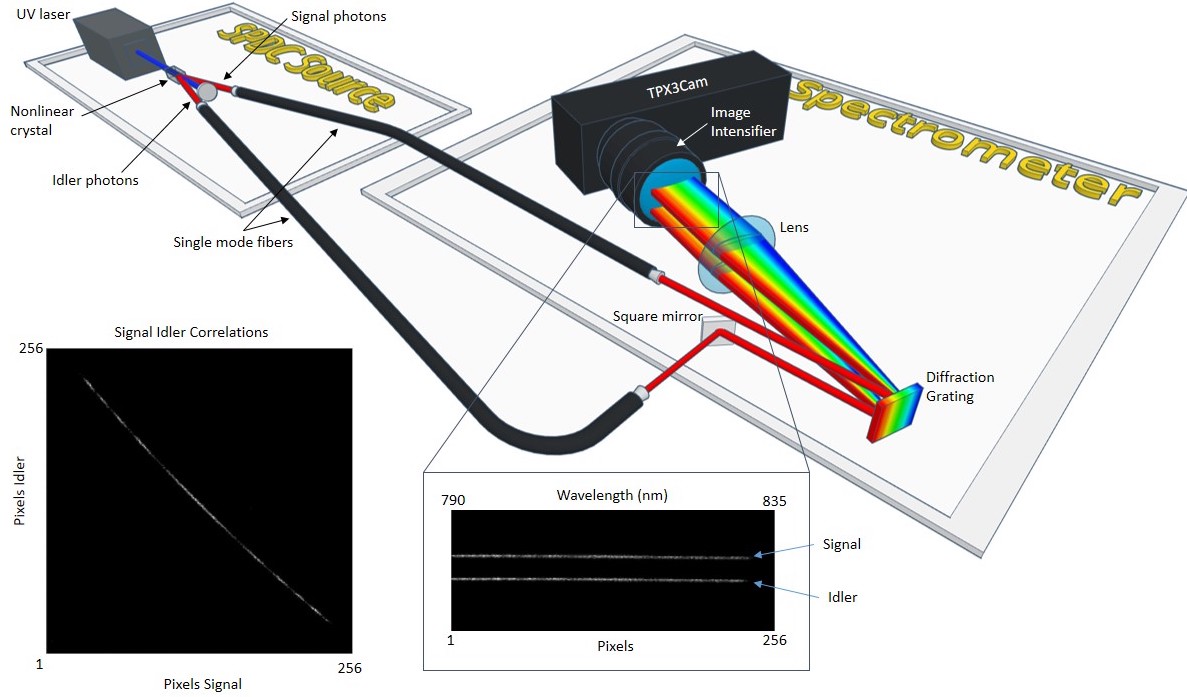}
\includegraphics[width=0.42\linewidth]{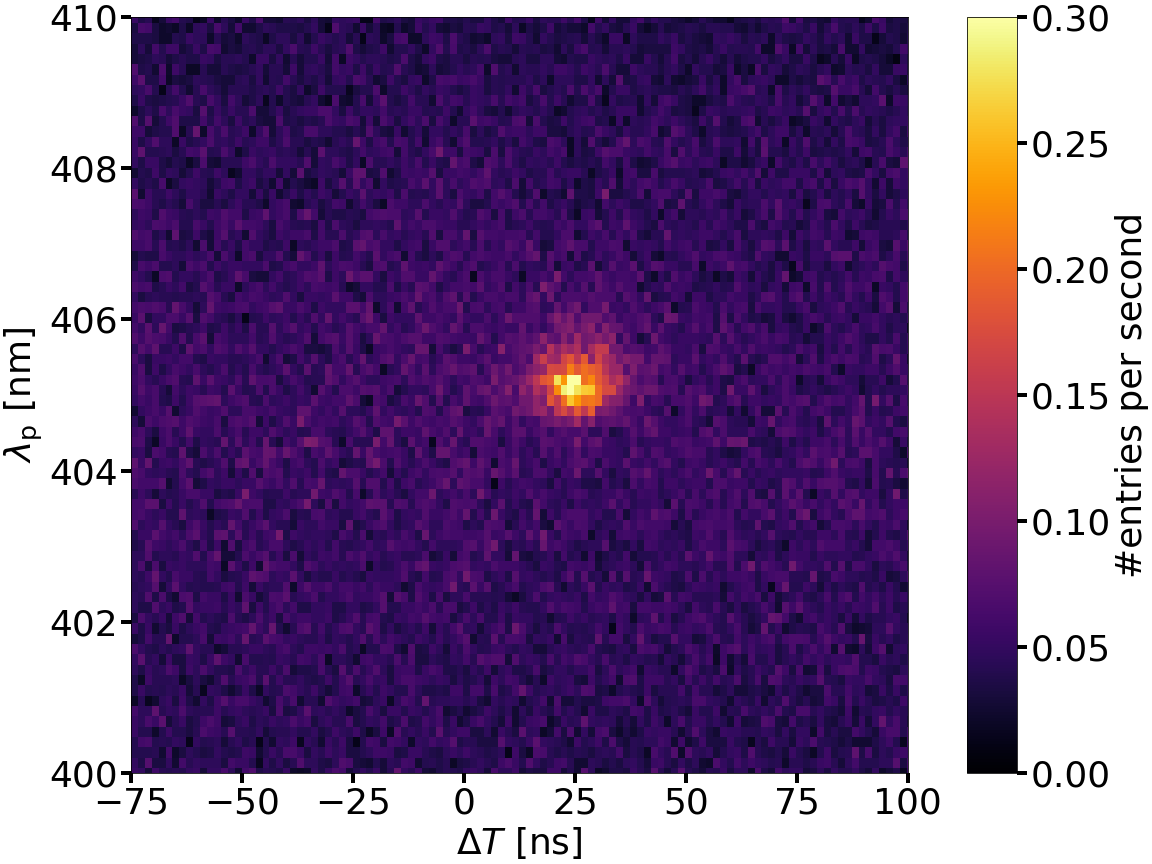}
\caption{Left: Demonstration of spectroscopic binning. The inset in the figure shows two stripes in the camera's raw data, which correspond to two photons from the spontaneous parametric down-conversion (SPDC) source analysed spectroscopically by the diffraction grating. The second insert shows the anti-correlation of the two photons' wavelengths due to the conservation of energy. Right: reconstructed wavelength of the pump photon and time difference between two time-stamped photons in the pair. The bright spot in the middle corresponds to the genuine photon pairs from the source.}
\label{fig:quantumtarget}
\end{center}
\end{figure}

We implemented the spectroscopic binning employing the fast camera and determined the temporal and spectroscopic resolutions for pairs of registered photons in the context of quantum target detection \cite{Zhang2020}.  The left part of Figure \ref{fig:quantumtarget} shows the setup used for the measurements. The inset in the figure shows two stripes in the camera's raw data, which correspond to two photons from the spontaneous parametric down-conversion (SPDC) source analysed spectroscopically by the diffraction grating. The second insert shows the anti-correlation of the two photons' wavelengths due to the conservation of energy. The right part of the figure shows the reconstructed wavelength of the pump photon and time difference between the two time-stamped photons in the pair \cite{Svihra2020}. The bright spot in the middle corresponds to the genuine photon pairs from the source. The distribution width is characteristic of the experimentally achievable spectral and temporal resolutions. Note that both were measured for a pair of photons, so the single photon resolutions should be adjusted (decreased) by the square root of two. This result is presented as a blue dot in Figure \ref{fig:uncertainty}.

\begin{figure}
\begin{center}
\includegraphics[width=0.95\linewidth]{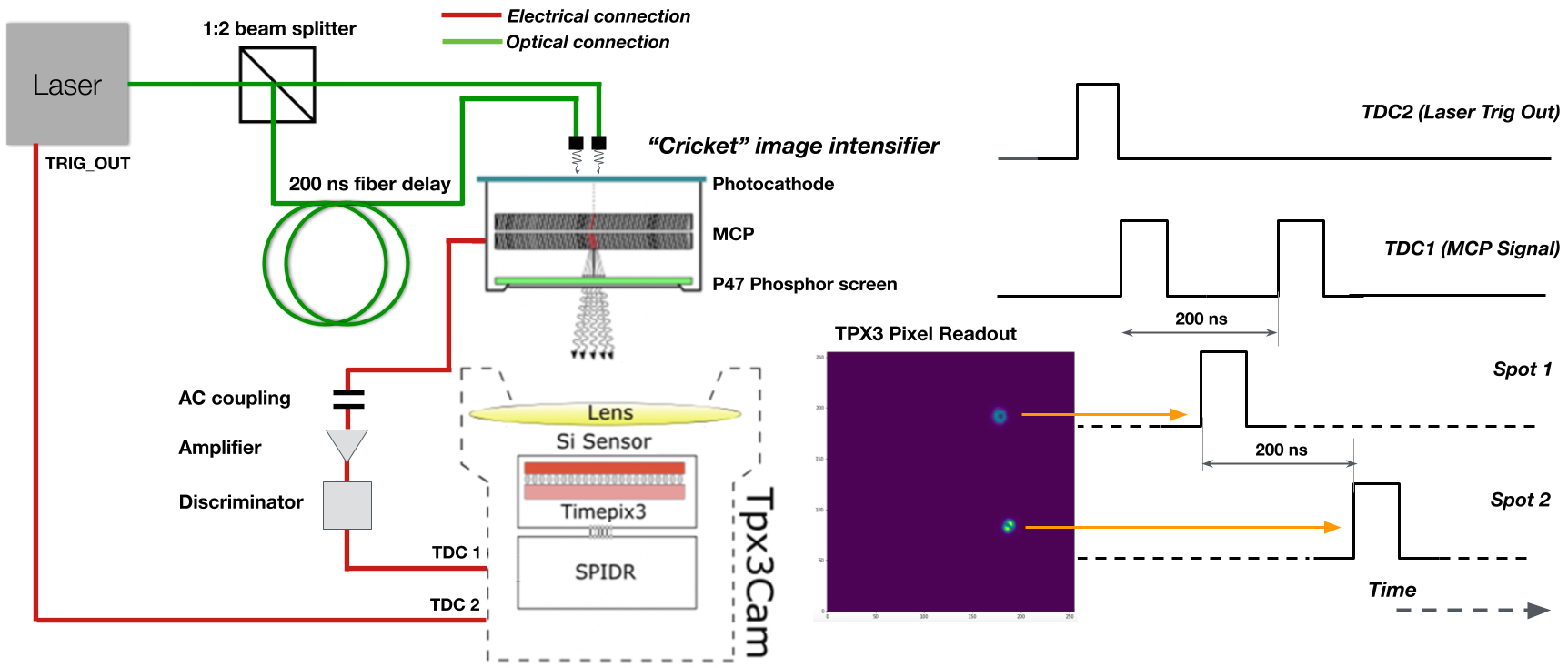}
\caption{Left: setup for the MCP time resolution measurements, see the text. Right: corresponding signal time diagram. Insert: Tpx3Cam pixel display with intensity of two laser spots.}
\label{fig:MCPoutput}
\end{center}
\end{figure}

The MCP inside the camera intensifier amplifies the single photon signals. 
The MCP output signal is very fast and can be used to improve the timing resolution, in principle, to the 30-ps level \cite{Debrah2020}. We tested this approach by employing an intensifier with MCP output connected to a fast 380 MHz ORTEC amplifier after a 1 MHz high-pass filter. The amplifier output was fed to a constant threshold LeCroy discriminator with its digital output time-stamped in the Tpx3Cam TDC with 0.26-ns granulation. The TDC is integrated with the Timepix3 readout, time-stamping the input signals in synchronization with the pixel data. The timing resolution was determined using a fast laser beam split into two paths, with one of them delayed by 200 ns as shown in Figure \ref{fig:MCPoutput}. This scheme produces two laser spots at the input of the intensifier and, therefore, two fast pulses at the MCP that then are time-stamped with the TDC. The right part of the figure shows the time diagram for signals registered by two channels of the camera TDC and Timepix3 pixels. The figure insert presents the Tpx3Cam pixel display with intensity of two laser spots. For the measurements described here we used the Photonis Cricket\texttrademark ~with the intensifier hi-QE-green photocathode. 

\begin{figure}
\begin{center}
\includegraphics[width=0.31\linewidth]{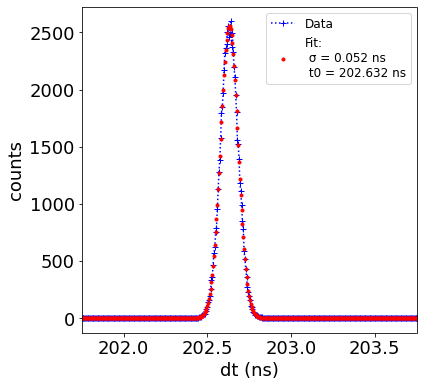}
\includegraphics[width=0.3\linewidth]{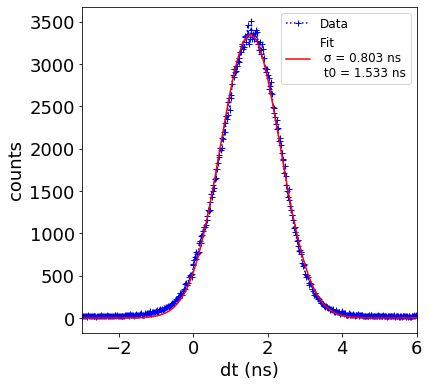}
\includegraphics[width=0.3\linewidth]{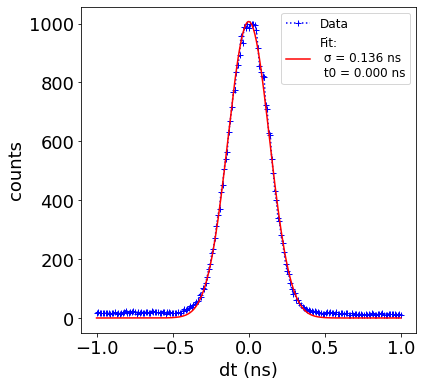}
\caption{Time resolution for three types of single photon detectors used for the benchmarking measurements. a) MCP readout of the intensified Tpx3Cam. Note that in this case it is not a single-photon regime; see the text; b) SPAD-based single photon counters; c) superconducting nanowire single photon counter. The time difference distributions are fit with a Gaussian function, with its sigma presented in the plots.}
\label{fig:timedifference}
\end{center}
\end{figure}

The time difference between the two pulses, about 202 ns, measured experimentally, is shown in Figure \ref{fig:timedifference} in the left graph. The distribution is well described with a Gaussian with sigma of 52 ps. In this case the data was acquired using a fast TDC (quTAG qutools) with 7-ps temporal resolution so not with the Tpx3Cam TDC, since the 260-ps binning of the latter would limit the measured resolution. To get the resolution per photon spot, this needs to be divided by square root of two corresponding to the spot resolution of 37 ps.
However, it should be noted that this measurement was not performed with single photons but rather for a signal of hundreds of photons emitted in a fast laser flash because the noise floor of the setup did not allow us to achieve single-photon sensitivity. This was most likely due to the noise performance of the intensifier power supply, which was integrated into the cricket assembly; work is in progress to improve it. Nevertheless, these results are encouraging and demonstrate the potential of the approach. The good timing performance of an MCP is a well-known fact, used, for example, in MCP-based photomultipliers (MCP-PMT) and other devices \cite{valerga2014, Vallerga2008, Tremsin2020}.

\subsection{Optical sensors with internal amplification}

A promising technique which allows excellent timing resolution is to replace the external intensifier with internal charge amplification inside the silicon sensor.
In certain conditions the n-p junction in silicon sensors can operate in the charge multiplication mode, providing internal amplification for the produced photoelectrons. The most common sensors where this can be implemented are LGAD (low gain avalanche device) \cite{Pellegrini2016, Giacomini2019, Cartiglia2020} and SPAD (single photon avalanche device)  \cite{Gasparini2017, Perenzoni2016, Lee2018} sensors. In the former, as their name implies, the gain is limited to factors of a few dozen and so does not produce large enough signals to ensure the single-photon sensitivity. This is true, at least, for the present implementations of prospective readout chips with fast time stamping such as Timepix3, discussed above but, likely, could be achieved for specialized monolithic implementations with improved noise performance. In SPAD devices, however, avalanche in the junction breaks it down, similar to the Geiger mode of the gaseous discharge, producing a fast pulse of big enough amplitude to operate with single-photon sensitivity.

These types of devices with internal amplification can have excellent timing resolution, which can be as good as 10 ps for single-channel devices. Multi-channel systems and imagers based on this technology have also started to appear on the market.  Various architectures are possible for SPAD designs, some integrating multiple cells into a single large-area device with photon counting capabilities \cite{Jiang2007} and some using individual SPAD cells. The former devices sometimes are referred to as silicon photomultipliers (SiPM). In the latter case the advances in the complementary metal-oxide-semiconductor (CMOS) technology are enabling integration of these devices into pixelated  sensors. This led to production of SPAD arrays capable of counting and time stamping single photons with resolution below 100~ps \cite{Gasparini2017, Morimoto2020}.

The photon detection efficiency (PDE) of SPAD devices can, in principle, be high in the range from 350 to 950 nm since they are made of silicon, which could perform with 90\% quantum efficiency, but so far the maximum achieved PDE is modest, around 30-40\% for multi-pixel devices. This is mostly due to the reduced fill-factor to implement protection from the inter-pixel crosstalk and constraints of junction implementation, which needs to be implanted at certain depth. The PDE of single-channel devices can be substantially higher, up to 60-70\%.

We benchmarked performance of commercial single photon counters based on this technology, namely a Laser Components USA Inc. COUNT\textregistered-50N, using the quTAG TDC and a SPDC single photon pair source which we used in our previous experiments \cite{Ianzano2020, Nomerotski2020}. Two photons in the source are produced at the same time in a down-conversion process so a simultaneous registration of these photons in two counters allows us to estimate their time resolution. Time difference between two photons from the source is shown in the central graph of Figure \ref{fig:timedifference}. Its timing resolution, about 570 ps per photon, is not the best one possible and is characteristic of the choice of available counters rather than an example of the best possible SPAD time resolution. 

\subsection{Superconducting sensors}

Superconducting nanowire single photon detectors (SNSPD) is an emerging quantum sensor technology, which employs narrow superconducting wires to detect single photons. A photon depositing its energy in the vicinity of the wire disrupts the superconductivity locally, inducing a voltage signal in the detection circuit  \cite{Divochiy2008, Zhu2020, Korzh2020}.

We employed a commercial multi-channel infrared-sensitive SNSPD system (SingleQuantum EOS) and paired it with the quTAG TDC module. We used the same SPDC setup as described in the previous subsection to characterize the detector channels' timing resolution. Time difference between two photons from the SPDC source is shown in the right graph of Figure \ref{fig:timedifference} with sigma of a Gaussian fit equal to 136 ps corresponding to resolution of 96 ps for a single photon. 

The superconducting nanowire detectors have excellent photon detection efficiency, in excess of 90\%. Recent improvements in the detector geometry and readout circuitry have yielded the demonstrations of three-picosecond timing resolution~\cite{Korzh2020} and small-scale pixel arrays \cite{Miyajima2018,Allmaras2020}.

\subsection{Streaking cameras}

\begin{figure}
\begin{center}
\includegraphics[width=0.9\linewidth]{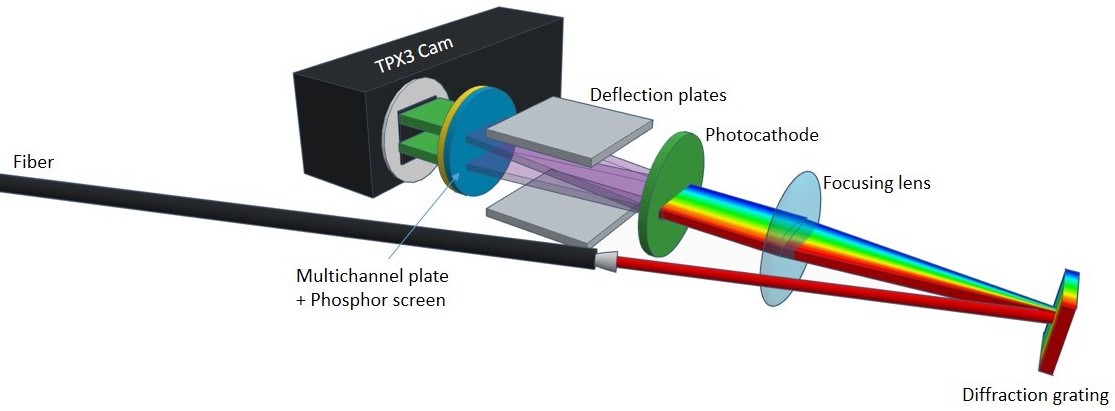}
\caption{A streaking tube, as shown in the figure, is a photo-sensitive vacuum device with a photocathode, a pair of deflecting electrodes, an MCP and a fast scintillating screen.  By applying a rapidly oscillating signal to the deflector, the photoelectrons are “streaked” onto the camera screen, effectively trading a temporal degree of freedom for a spatial one.}
\label{fig:streaking}
\end{center}
\end{figure}

Streaking cameras offer a different method of time measurement by replacing the direct time determination with a position measurement which can be interpreted as time.
The layout of a streaking camera is schematically shown in Figure \ref{fig:streaking}. Photons are converted into photoelectrons on a photocathode. Then the photoelectrons are spread in the vertical direction by a variable electric field before they are amplified with an MCP and produce flashes on a fast scintillating screen. The precise timing can be determined by the vertical coordinate of registered photons. In one possible implementation, a standalone streak tube could be attached to a fast camera like Tpx3Cam.

Thus, the streaking tube is a photo-sensitive vacuum device with a photocathode, an MCP, and a fast scintillating screen. This arrangement is quite similar to the intensifier described in the previous subsection, but in addition, it also has a pair of deflecting electrodes.  By applying a rapidly oscillating signal to the deflector, the photoelectrons are “streaked” onto the screen, effectively trading a temporal degree of freedom for a spatial one. The streak tube could be read out with the same fast camera, Tpx3Cam, with nanosecond time resolution and 80-Mpix/s throughput.

The data will look similar to the inset in Figure \ref{fig:quantumtarget}, but the two stripes will be smeared in the vertical direction according to their timing. This will make use of the spare space in the camera sensor that was not used in our previous work \cite{Zhang2020}.

The streaking schemes are known to be able to achieve temporal resolution down to 1 ps \cite{Korobkin1969, Chollet2008, Howorth2016}. To date this appears to be the most precise practical technique to time-stamp single photons. According to Figure \ref{fig:uncertainty} this kind of time resolution would require only $\sim$1 nm spectral binning. The drawback of the technique is the need to deflect macroscopic trajectories of charged particles, photoelectrons,  in the vacuum and, therefore, the need for a photocathode. This would limit the achievable detection efficiency to about 30\%.

\section{First Experiments}

\begin{figure}
\begin{center}
\includegraphics[width=0.4\linewidth]{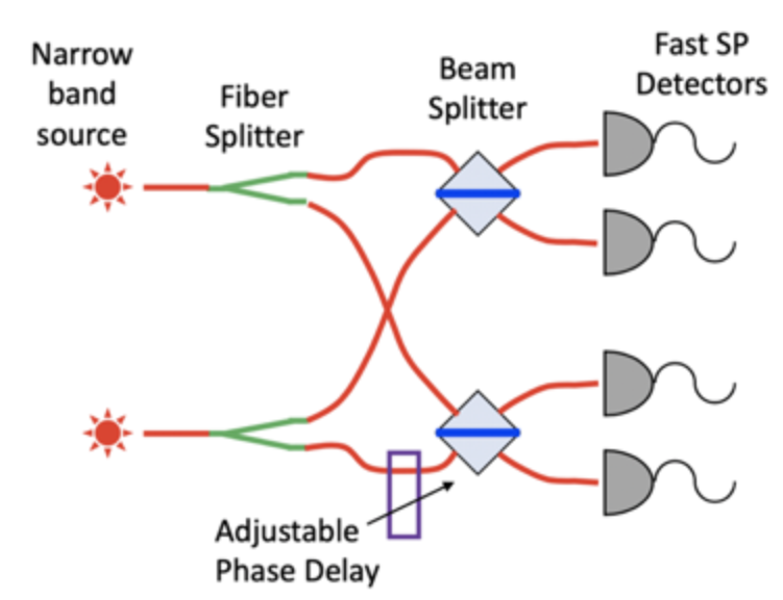}
\includegraphics[width=0.59\linewidth]{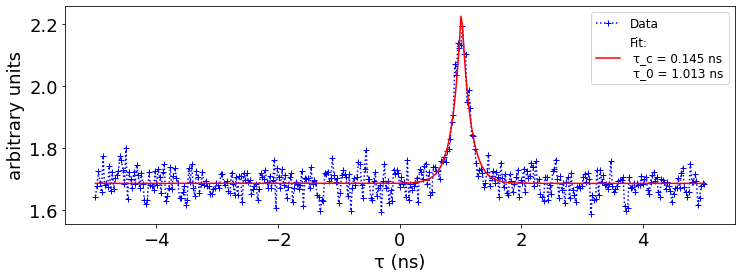}
\caption{Left: Diagram of first experiments implemented with argon lamps as thermal sources and fiber-coupled beam-splitters. Right: Example of the HBT peak fit with a Lorentzian function with decay time of $0.145 \pm 0.01$ ns.}
\label{fig:HBT}
\end{center}
\end{figure}

The first experiments to demonstrate the two-photon amplitude interference for the astrometric applications were implemented with argon lamps as quasi-thermal sources and fiber-coupled beam-splitters. The setup is shown schematically in the left part of Figure \ref{fig:HBT}. 

We used a spectral line at $794.818 \pm 0.001$ nm in a low-pressure argon lamp as a narrow-band source of thermal photons. The lamp was used with a $794.9 \pm 1$ nm filter. It is expected that the natural lifetime of the line is $\sim1$ ns, corresponding to the natural bandwidth of GHz, and that the Doppler broadening due to the atomic thermal velocities does not affect it in a considerable way. The source coherence time was evaluated by observing the Hanbury Brown \& Twiss (HBT) effect of intensity interference.
The right part of Figure \ref{fig:HBT} gives an example of the HBT peak 
fit with a Lorentzian function with decay time of $0.145 \pm 0.01$ ns. 

As shown in the figure, we used four SNSPDs (SingleQuantum EOS), and the plot in Figure \ref{fig:HBT} represents a correlation between photons detected at a specific pair of detectors. However, we observed similar peaks for all six combinations of two detectors. 
The width of the peak has a contribution from the system time resolution, presented in the previous section for the SNSPD counters in Figure \ref{fig:timedifference}.

Work is in progress to measure effects associated with change of the phase between the two thermal sources, which would correspond to their relative displacement in space. We are also planning on-sky measurements with bright stars to further validate the two-photon interference technique.

\section{Conclusions}

We have reviewed the theory motivating a novel optical interferometer with significantly improved astrometric precision first proposed in our previous work \cite{stankus2020} and provided a detailed discussion of the requirements for its effective operation. Utilizing quantum mechanical two-photon interference effects from two sky sources, we demonstrate that the traditionally required connective path between optical telescopes can be bypassed, allowing for arbitrarily extended baseline separations and dramatically improved astrometric precision. 

Through a quantum optics framework and an idealized implementation of the two-photon amplitude interferometer, we detail the instrument’s efficacy. Pair coincidence measurements at each detection site can be used to identify the fringe rate due to the Earth rotation, which is directly related to the relative astrometry of the two sources. Following a Fisher matrix formulation, we show that with reasonable assumptions for the instrument and observed astronomical sources, a single night’s observation of two bright stars could lead to opening angle precision on the order of $10\,\mu$as.

We then move to a discussion of the spectroscopic and temporal performance requirements for effective operation of the instrument. Two-photon amplitude interferometry requires that the photons be indistinguishable imposing demanding spectral and temporal resolution restrictions. The Heisenberg limit can be used to set the scale for the resolution that must be achieved to satisfy the indistinguishability condition. Finer spectral resolution relaxes the temporal resolution requirements and vice versa. 

Existing manufacturing capabilities allow for first-order diffraction gratings with spectral resolution around 0.05 nm, achieving a spectral resolution that, if integrated with sufficiently sensitive imaging, is satisfactory. The spectral resolution can be further improved upon with existing technologies, such as Echelle gratings.
We also discussed several sensor technologies which show merit as candidates capable of achieving the necessary timing resolution at the picosecond scale. 

We end with a discussion of preliminary results from first experiments as important first steps in establishing the baseline coherence time of the observed HBT effect and the timing performance of the single photon detectors. Proof-of-concept experiments with two quasi-thermal sources and a variable phase delay are ongoing to demonstrate the predicted phase-dependent two-photon interference effects, eventually paving the way for testing with sky sources. 

\acknowledgments 
 
We thank Thomas Jennewein, Emil Khabiboulline and Ning Bao for inspiring discussions. This work was supported by the U.S. Department of Energy QuantISED award and BNL LDRD grant 19-30. S.A, A.P. and J.S. acknowledge support under the Science Undergraduate Laboratory Internships (SULI) Program  by the U.S. Department of Energy.

\bibliography{report} 
\bibliographystyle{spiebib} 

\end{document}